\documentclass[12pt]{article}
\usepackage[dvips]{graphicx}
\usepackage[dvips]{graphics}

\newcommand{\newc}{\newcommand}
\newc{\lra}{\leftrightarrow}
\newc{\beq}{\begin{equation}}
\newc{\eeq}{\end{equation}}
\newc{\barr}{\begin{eqnarray}}
\newc{\earr}{\end{eqnarray}}

\title{Direct dark matter
search by observing electrons produced in neutralino-nucleus
collisions}

\author{Ch.~C.~Moustakidis$^1$\footnote{\texttt{e-mail:\, moustaki\,@\,auth.gr}}\,,
J.~D.~Vergados$^2$\footnote{\texttt{e-mail:\,
Vergados\,@\,cc.uoi.gr}}\,, and
H.~Ejiri$^3$\footnote{\texttt{e-mail:\,
ejiri\,@\,rcnp.osaka-u.ac.jp}}
\\
{\it  $^{1}$Department of Theoretical Physics,} \\
{\it Aristotle University of Thessaloniki,} \\
{\it 54124 Thessaloniki Greece}\\
{\it $^{2}$University of Ioannina, 45110 Ioannina, Greece} \\
{\it $^{3}$ NS, International Christian University,}\\
{\it Osawa, Mitaka, Tokyo, 181-8585, Japan}\\
}

\date{}

\begin{document}

\maketitle

\begin{abstract}
Exotic dark matter and dark energy together seem to dominate in
the Universe.
Supersymmetry naturally provides a candidate for the dark matter
constituents via the lightest supersymmetric particle (LSP). The
most important process for directly detecting dark matter is the
LSP-nucleus elastic scattering by measuring the energy of the
recoiling nucleus. In the present work we explore a novel process,
which has definite experimental advantages,
 that is the detection of the dark
matter constituents by observing the low energy ionization
electrons. These electrons, which are produced during the
LSP-nucleus collision, may be observed separately or in
coincidence with the recoiling nuclei.
 We develop the formalism and apply it in calculating the ratio of the ionization rate to the
 nuclear recoil rate in a variety of atoms
including $^{20}$Ne, $^{40}$Ar, $^{76}$Ge, $^{78}$Kr and
$^{132}$Xe employing realistic Hartree-Fock electron wave
functions.
  The obtained ratios are essentially
independent of all parameters of supersymmetry except the
neutralino mass, but they crucially depend on the electron energy
cut off. These ratios  per electron tend to
 increase with the atomic number and can be as high as $10\%$. Based on our results it is both interesting and
realistic to detect the LSP by measuring the ionization electrons
following the-LSP nuclear collisions.
\end{abstract}

\noindent Keywords: LSP, Direct neutralino search, ionization
electrons, electron detection,
 dark matter, WIMP.\\
PACS : 95.35+d, 12.60.Jv.\\\\\

\section{Introduction}
From the rotational curves of objects outside the luminous
galaxies it has been known
 for some time that there must be dark (non luminous) matter in the Universe.
Furthermore the combined MAXIMA-1 \cite{MAXIMA-1}, BOOMERANG
\cite{BOOMERANG},
 DASI \cite{DASI} and COBE/DMR \cite{COBE} Cosmic Microwave Background (CMB)
observations as well as the high precision recent WMAP data
\cite{SPERGEL}-\cite{SDSS} imply that the Universe is flat
\cite{flat01} and it is dominated by dark matter and dark energy
(cosmological constant). Crudely speaking  one has:
$$\Omega_b=0.05, \qquad \Omega _{CDM}= 0.30, \qquad \Omega_{\Lambda}= 0.65$$
for the baryonic, dark matter and dark energy fractions,
respectively.
 It is also known that
the non exotic component cannot exceed $40\%$ of the CDM ~\cite
{Benne}. So there is plenty of room for the exotic WIMPs (Weakly
Interacting
Massive Particles).\\
  Many experiments are currently under way aiming at the direct detection
 of WIMPs.  In fact the DAMA experiment ~\cite {BERNA2} has claimed the
  observation of  such events, which with better statistics
   have subsequently been interpreted
as a modulation signal \cite{BERNA1}. These data, however, if
interpreted as due to the scalar interaction, are not consistent
with other recent experiments, see e.g. EDELWEISS \cite{EDELWEISS}
and CDMS \cite{CDMS}.

 Supersymmetry naturally provides candidates for the dark matter constituents
\cite{Jung},\cite{GOODWIT}-\cite{ELLROSZ}.
 In the most favored scenario of supersymmetry the
lightest supersymmetric particle (LSP) can be simply described as
a Majorana fermion, a linear combination of the neutral components
of the gauginos and higgsinos
\cite{Jung},\cite{GOODWIT}-\cite{WELLS}.

 The event rates, however, are expected to be quite low and the nuclear
recoil energies are extremely small. Thus one has to try to reduce
the background to the lowest possible level and to understand
possible origins of  backgrounds to be corrected for \cite{VerEj}.
Anyway one has to search for characteristic signatures associated
with the LSP. Such are:
\begin{itemize}
\item The time dependence of the event rates, which is due to the
motion of the Earth (modulation effect). Unfortunately, however,
the modulation effect is small, $\le2\%$, and for heavy targets
the relevant sign depends on the LSP mass. So one cannot
accurately predict even when  the maximum occurs. Furthermore one
should keep in mind that there are seasonal effects on the
background as well. \item The correlation of the observed rates of
directionally sensitive experiments with the motion of the sun
\cite{SLSKA,DIREXP,JDV03}. In this case one has large asymmetries
and larger relative modulation amplitudes, with characteristic
directional dependence. These are, indeed, very interesting
signatures, but the expected event rates are suppressed by a
factor of at least $4\pi$ relative to the standard rates.
Furthermore, at present, the  experiments are able to measure the
direction of nuclear recoil, but it will be extremely difficult to
determine the sense of direction \cite{TPCSpooner}. \item
Inelastic excitations  to low lying excited nuclear states have
also been considered \cite{VQD03,EFO93}. In this case one can
overcome the difficulties encountered in the recoil experiments
(energy threshold, quenching factors etc) by detecting the
de-excitation $\gamma$-rays. The predicted branching ratios, due
to the spin induced DM cross sections,
 in favorable circumstances can be as high as $10\%$, which is a very encouraging result.
\item Detecting the dark matter constituents by observing the low
energy electrons, which follow the ionization of the atom during
the LSP-nucleus collision \cite{VerEj}.
\end{itemize}
 The last possibility may be realized with the
technology of gaseous TPC detectors \cite{GV03}-\cite{GRRC}. In
fact the WIMP-nucleus scattering leads: (i) to nuclear recoil
without atomic excitations and (ii) to nuclear recoil with atomic
excitations. So far most CDM searches have been made by the
inclusive processes, (i) and (ii), employing solid detectors. We
propose that the produced electrons in atomic excitation (ii)
should be studied by exclusive measurements. In a previous
calculation \cite{VerEj} we have estimated the ratio of the
ionization rate (per electron) by considering the simple target
$^{20}$Ne
assuming hydrogenic electron wave functions.\\
 In the present paper
we extend these calculations by considering a number of
experimentally interesting targets. To this end realistic
Hartree-Fock electron wave functions have been employed. The
observed ratio of the electron rate (ii) for one electron per
atom, to be denoted below as per electron,
 divided by the standard
nuclear recoil rate (i), depends, of course,  on the low energy
cut off for both electrons (tenths of keV) and nuclear recoils
(10-20 keV) as well as the quenching factor for recoils. They can
reach
It is found to increase as the atomic number and it can get as
high as $10\%$ for a reasonable choice of these parameters.  The
experimentally interesting quantity the ratio per atom, meaning
that one considers all electrons in the atom, obtained from the
previous one by multiplying it with $Z$. This ratio for medium and
heavy nuclei may even become larger than unity. Thus this method
can be widely used for any
target nuclei with adequate electron detection capability.\\
 The present paper is organized as follows: The theoretical framework is presented in Sec. \ref{sec2}, the main results are shown in
 Sec. \ref{sec3} and a brief summary is given in Sec. \ref{sec4}.
\section{The essential ingredients of our calculation}
\label{sec2}
 The electron ionization due to the LSP collisions can take place via two mechanisms:
 \begin{enumerate}
 \item The electrons are ejected during the LSP-nucleus collisions
 via the standard EM interactions.
 \item The electrons are ejected due to the LSP-electron interaction.
 This interaction is obtained in a
fashion analogous to the LSP-quark interaction. In the case of the
Z and Higgs exchange one replaces the relevant quark couplings
\cite{JDV96} by the electron couplings. In the s-fermion mediated
mechanism the s-quarks are replaced by s-electrons. In fact this
mechanism has recently been proposed as providing a possibility of
LSP detection
 in electron accelerators \cite{HNNS05}. In the present case,
 however, one can get reasonably high electron
energies, of a few hundreds of electron volts, only if one
exploits electrons bound in nuclei. Then the kinematics is similar
to that of the previous mechanism.
\end{enumerate}
 Since the branching ratios of the second case depend on the
 SUSY parameter space and one does not enjoy the effect
of coherence arising from the scalar interaction, we will
concentrate here in the first mechanism.
\subsection{The formalism}
The differential cross section for the LSP nucleus scattering
leading to the emission of electrons in the case of
non-relativistic neutralino takes the form \cite{VerEj}
\begin{eqnarray}
d\sigma({\bf k})& &=\frac{1}{\upsilon}\frac{m_e}{E_e}|M|^2
\frac{d{\bf q}}{(2\pi)^3} \frac{d{\bf k}}{(2 \pi)^3} (2 \pi)^3
\frac {1}{2(2l+1)}
 \sum_{n \ell m}p_{n \ell} \times \\ \nonumber
& & \left[ \tilde{\phi}_{n \ell m}({\bf k}) \right]^2 2 \pi
\delta\left(T_{\chi}+\epsilon_{nl}-T-\frac{q^2}{2m_A}-\frac{({\bf
p}_{\chi}-{\bf k}-{\bf q})^2}{2m_{\chi}} \right) \label{CS-1}
\end{eqnarray}
where $\upsilon$, $T_{\chi}$ and ${\bf p}_{\chi}$ are the oncoming
LSP velocity, energy and momentum distribution, while ${\bf q}$ is
the momentum transferred to the nucleus. $M$ is the invariant
amplitude, known from the standard neutralino nucleus cross
section, $T$ and ${\bf k}$ are the kinetic energy and the momentum
of the outgoing electron and $\epsilon_{nl}$ is the energy of the
initial electron, which is, of course, negative.
$\tilde{\phi}_{nlm}({\bf k})$ is the Fourier transform of the
bound electron wave function, i.e., its wave function in momentum
space. $p_{nl}$ is the probability of finding the electron in the
$n$, $l$ orbit. In the expression above and in what follows our
normalization will consist of one electron per atom, to be
compared with the cross section per nucleus of the standard
experiments.

In order to avoid any complications arising from questions
regarding the allowed SUSY parameter space, we will present our
results normalized to the standard neutralino-nucleus cross
section. The obtained branching ratios are essentially independent
of all parameters of supersymmetry except the neutralino mass.

With these ingredients we find that the ratio of the cross section
with ionization divided by that of the standard neutralino-nucleus
elastic scattering, nuclear recoil experiments (nrec), takes the
form
\begin{eqnarray}
\frac{d\sigma(T)}{\sigma_{nrec}}&=& \frac{1}{4} \sum_{n \ell} p_{n
\ell}
|\tilde{\phi}_{n \ell}(\sqrt{2m_eT})|^2 m_e \sqrt{2m_eT} \nonumber \\
&\times& \frac{\int_{-1}^{1} d\xi_1 \int_{\xi_L}^{1} d\xi K
\frac{(\xi+\Lambda)^2}{\Lambda} [F(\mu_r \upsilon
(\xi+\Lambda))]^2} {\int_{0}^{1} 2\xi d\xi [F(2m_r \upsilon
\xi)]^2}  dT \label{CS-2}
\end{eqnarray}
where $\mu_r$ the LSP-nucleus reduced mass and also
\begin{eqnarray}
\Lambda&=&\sqrt{\xi^2-\xi_L^2}
\nonumber \\
\xi_L&=&\sqrt{\frac{m_{\chi}}{\mu_r}[1+\frac{1}{K^2}(\frac{T-\epsilon_{n
\ell}}{T_{\chi}}-1)]} \label{CS-3}
\end{eqnarray}
with
\begin{eqnarray}
{\bf K}&=&\frac{{\bf p}_{\chi}-{\bf k}}{p_{\chi}}, \quad
K=\frac{\sqrt{p_{\chi}^2+k^2-2kp_{\chi}\xi_1}}{p_{\chi}} \nonumber\\
\xi_1&=&\hat{p}_{\chi}\cdot \hat{k}, \quad \xi=\hat{q}\cdot
\hat{K} \label{CS-4}
\end{eqnarray}
$2\frac{\mu_r}{m_{\chi}}p_{\chi}\xi=2\mu_r\upsilon \xi$ is the
momentum $q$ transferred to the nucleus and $F(q)$ is the nuclear
form factor. For a given LSP energy, the outgoing electron energy
lies in the range given by:
\begin{equation}
0\leq T \leq \frac{\mu_r}{m_{\chi}}T_{\chi}+\epsilon_{n \ell}.
\label{linitT}
\end{equation}
Since the momentum of the outgoing electron is much smaller than
the momentum of the oncoming neutralino, i.e., $K\approx 1$, the
integration over $\xi_1$ can be trivially performed. Furthermore,
if the effect of the nuclear form factor can be neglected, the
integration over $\xi$ can be performed analytically. Thus we get
\begin{eqnarray}
\frac{d\sigma(T)}{\sigma_{nrec}}& &=\frac{1}{2} \sum_{n \ell} p_{n
\ell}
|\tilde{\phi}_{n \ell}(\sqrt{2m_eT})|^2 m_e \sqrt{2m_eT} \times \nonumber \\
&
&\left[1-\left(\frac{m_{\chi}}{\mu_r}\frac{(T-\epsilon_{nl})}{T_{\chi}}\right)+
\sqrt{1-\left(\frac{m_{\chi}}{\mu_r}\frac{(T-\epsilon_{n
\ell})}{T_{\chi}}\right)}\right]
 dT \label{CS-5}
\end{eqnarray}
Otherwise the angular integrations can only be done numerically.
 Finally one must integrate numerically the above expression over the
electron spectrum to obtain the total cross section as a function
of the neutralino energy (velocity). The folding with the LSP
velocity will be done at the level of the event rates (see sec.
\ref{convolution})
\subsection{The role of the nuclear form factor}
The nuclear form factor $F(u)$ entering in Eq. (\ref{CS-2}) has
the general form
\begin{equation}
F(u)=\frac{Z}{A}F_Z(u)+\frac{N}{A}F_N(u) \label{FF-1}
\end{equation}
The proton and neutron form factors ($F_Z(u)$ and $F_N(u)$
respectively) are calculated by the Fourier transform of the
proton density distribution $\rho_Z({\bf r})$ and neutron density
distribution $\rho_N({\bf r})$ respectively. In the present work
$\rho_Z({\bf r})$ and $\rho_N( {\bf r})$ are constructed  using
harmonic oscillator wave functions. More specifically the
densities are defined as
\begin{equation}
\rho_{Z,N}({\bf r})=\sum_{i}\phi_i^{*}({\bf r}) \phi_i({\bf r})
\label{HOWF-1}
\end{equation}
where the sum run all over the proton ( neutron) orbitals
$\phi_i({\bf r})$.  In general some of the nuclei considered in
this work may be deformed. Thus, strictly speaking, the densities
depend not only on the radial coordinate $r$ but on the angle
$\theta$ as well. In the present work, however, we are mainly
interested in branching ratios, which depend very little on the
nuclear form factor.  Thus to a good approximation the nuclei can
be treated as spherical. Thus in the case of the heaviest nuclear
system considered in this work, namely $^{132}$Xe, for a threshold
energy of $E_{th}=0.2$ keV the effect of the form factor is a
reduction of the branching ratio by less than $7\%$. For this
reason we have included only one stable isotope, namely the
 most abundant one, which is
indicated on the figures.\\
In view of the above  the  form factor is given by the relation:
\begin{equation}
F_{Z,N}(q)=\int e^{i{\bf q}{\bf r}} \rho_{Z,N}( r) d{\bf r}
\label{FF-2}
\end{equation}
The nuclear form factor of $^{20}$Ne is taken from Ref.
\cite{Divari-00}, where the nuclear wave functions were obtained
by shell model calculations using the Wildenthal interaction. The
nuclear form factors of the heavier nuclei were for simplicity
calculated by using spherical harmonic oscillator wave functions.
This is a good approximation since, as it has already been
mentioned,
 the ratio of the two processes we is insensitive to the
 details of the nuclear wave functions.
\subsection{The bound electron wave functions}
In the present work we consider very accurate spin-independent
atomic wave functions obtained by  Bunge {\it et al}
\cite{Bunge-93}, by applying the Roothaan-Hartree-Fock method
(RHF) to calculate analytical self-consistent-field atomic wave
function. In principle one should employ the relativistic electron
wave functions, especially for the inner shell electrons of large
Z atoms. Anyway for our purposes the above wave functions are
adequate, since the inner shell electrons contribute only a small
fraction to the rate. In this approach the radial atomic orbitals
$R_{n \ell}$ are expanded as a finite superposition of primitive
radial functions
\begin{equation}
R_{n \ell}(r)=\sum_j C_{jn \ell} S_{j \ell}(r) \label{RHF-1}
\end{equation}
where the normalized primitive basis $S_{j \ell}(r)$ is taken as a
Slater-type orbital set,
\begin{equation}
S_{j \ell}(r)=N_{j \ell} r^{n_{j \ell}-1}e^{-Z_{j \ell}r}
\label{RHF-2}
\end{equation}
where the normalization factor $N_{j \ell}$ is given by
\begin{equation}
N_{j \ell}=(2Z_{j \ell})^{(n_{j \ell}+1/2)}/[(2n_{j \ell})!]^{1/2}
\label{Norm-1}
\end{equation}
and $n_{j \ell}$ is the principal quantum number, $Z_{j \ell}$ is
the orbital exponent, and $\ell$ is the azimuthal quantum number.

The atomic wave functions in the momentum space are related to the
coordinate wave functions by
\begin{equation}
\tilde{\phi}_{n \ell m}({\bf k})=\frac{1}{(2\pi)^{3/2}}\int e^{-i
{\bf k} {\bf r}}\phi_{n \ell m}({\bf r}) d{\bf r} \label{WFM-1}
\end{equation}
where $n \ell m$ denote the usual quantum numbers characterizing
atomic states. The radial momentum-wave function defined as
\begin{equation}
\tilde{\phi}_{n \ell m}({\bf k})=(-i)^{ \ell} \tilde{R}_{n
\ell}(k)Y_{ \ell m}(\Omega_k) \label{WFM-2}
\end{equation}
is related to the radial wave function in coordinate space through
(see also \cite{Gounaris-04})
\begin{equation}
\tilde{R}_{n \ell}(k)=\sqrt{\frac{2}{\pi}} \int_{o}^{\infty}
r^2R_{nl}(r)j_{ \ell}(kr) dr \label{WFM-3}
\end{equation}
where $j_l(kr)$ is a spherical Bessel function. The radial wave
functions in momentum space $\tilde{R}_{n \ell}(k)$ are written as
\begin{equation}
\tilde{R}_{n \ell}(k)=\sum_j C_{jn \ell} \tilde{S}_{j \ell }(k)
\label{WFM-4}
\end{equation}
in term of the RHF functions $\tilde{S}_{j \ell}(k) $ in momentum
space, related to $S_{j \ell}(r)$ through
\begin{equation}
\tilde{S}_{j \ell}(k)=\sqrt{\frac{2}{\pi}} \int_{0}^{\infty}
r^2S_{j \ell}(r) j_{ \ell}(kr) d r \label{WFM-5}
\end{equation}
The needed binding energies were taken from \cite{el-b-en}.
\subsection{Folding with the LSP velocity distribution}
\label{convolution}

 The above results were derived assuming a definite LSP velocity.
 In practice, however, the
 LSP obeys a velocity distribution. The actual velocity
 distribution may be complicated, i.e. asymmetric and
varying from one point to another. In
 our vicinity it is commonly assumed that LSP obeys a simple
 Maxwell-Boltzmann velocity
distribution with respect to the galactic center \cite{JDV03},
namely
\begin{equation}
f(v)=\frac{1}{(\upsilon_0 \sqrt{\pi})^3}e^{-(v^2/\upsilon_0^2)}
\label{MB-1}
\end{equation}
with $\upsilon_0=220$Km/s. This velocity distribution does not go
to zero at some finite velocity. So on this one imposes by hand an
upper velocity bound (escape velocity), $\upsilon_{esc}=2.84
\upsilon_0$. One then must transform this distribution to the lab
frame, ${\bf \upsilon} \rightarrow {\bf \upsilon}+ {\bf
\upsilon}_0$, since $\upsilon_0$ is also the velocity of the sun
around the center of the galaxy. We will not be concerned here
with the motion of the Earth around the sun, i.e., the modulation
effect \cite{JDV03}

Folding both the numerator and denominator of Eq. (\ref{CS-2})with
the LSP velocity distribution, after multiplying each with the LSP
flux $$\frac{\rho(0)}{m_{\chi}}\frac{m}{A m_p}\upsilon$$ we obtain
the differential ratio $\frac{1}{R}\frac{dR_e}{dT}$, with $R_e$
the rate for the ionization, on the form
\begin{eqnarray}
\frac{1}{R}\frac{dR_e}{dT}&=& \frac{d\sigma(T)}{\sigma_{nrec}}=
 \sum_{n \ell}p_{n \ell} |\tilde{\phi}_{n \ell}(\sqrt{2m_eT})|^2
m_e \sqrt{2m_eT}
 \nonumber\\
& & \times \frac{\int_{\upsilon_{min}}^{\upsilon_{max}} N
\upsilon^2 e^{-\upsilon^2/\upsilon_0^2}
\sinh(2\upsilon/\upsilon_0) d\upsilon}
{\int_{\upsilon{\prime}_{min}}^{\upsilon_{max}} D \upsilon^2
e^{-\upsilon^2/\upsilon_0^2} \sinh(2\upsilon/\upsilon_0)
d\upsilon}  \ dT
 \label{ratio-1}
\end{eqnarray}
where the numerator factor $N$ and the denominator factor $D$ are
given respectively
\begin{equation}
N=\frac{2}{a_1^2}\int_{a_2}^{a_3} q_b F^2(q_b) dq_b, \quad q_b=qb
\label{N-1}
\end{equation}
\begin{equation}
D=\frac{2}{a_1^2}\int_{0}^{a_1} q_b F^2(q_b) dq_b, \quad q_b=qb
 \label{D-1}
\end{equation}
where
\begin{equation}
a_1=\frac{2\mu_r \upsilon b}{\hbar c^2}, \qquad a_2=\frac{\mu_r
\upsilon b}{\hbar c^2} \xi_L, \qquad a_3=\frac{\mu_r \upsilon
b}{\hbar c^2}(1+\sqrt{1-\xi_L})
 \label{a123}
\end{equation}
and $b$ is the width of the harmonic oscillator potential.

In Eq. (\ref{ratio-1}) $\upsilon_{max}$ is the escape velocity,
while $\upsilon_{min}$ is given by Eq. (\ref{linitT}), i.e. \beq
\upsilon_{min}=\sqrt{ \frac{ 2 \left( T-\epsilon_{n \ell}
\right)}{\mu_r} } \label{vmin} \eeq while for the nuclear recoils:
\beq \upsilon{\prime}_{min}=\frac{ \sqrt{ 2 A m_p Q_{th}}}{2
\mu_r} \label{vminprime} \eeq
 It is clear that the fraction in Eq. (\ref{ratio-1}) is a decreasing function of the outgoing electron energy $T$ as well
as the binding energy $-\epsilon_{n\ell} $ of the initial
electrons.  For $T$ in the sub-keV region, where the differential
rate becomes sizable,  $v_{min}$ is determined essentially by the
energy of the bound electron. This restriction on the velocity
distribution is a source of suppression of the rate associated
with the inner shell electrons in the case of heavy targets,
 in particular, the $1s$ electrons. In fact for Xe for $T\cong 0$ this fraction reaches the maximum values of
$0.13,0.75,0.76$ and $0.95$ for 1s, 2s, 2p and 3s respectively.
For lighter systems this kinematical suppression is not important
even for the $1s$ orbits.
 Further suppression comes, of course, from the behavior of the bound electron wave function.
\section{Results and Discussion}
\label{sec3}
\begin{figure}
 \centering
 \includegraphics[height=8.0cm,width=8.0cm]{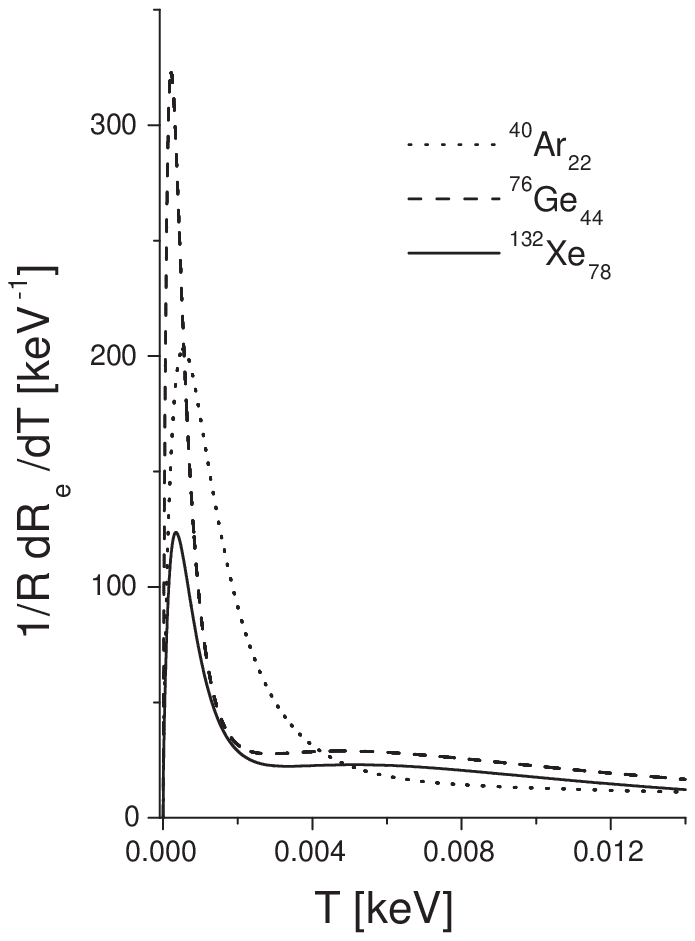}
  \caption{The differential rate for ionization electrons, divided by the total rate associated
  with the nuclear recoils, as a function of the electron energy $T$ (in keV) for
  various atoms. The results exhibited were obtained for
  a typical LSP mass $m_{\chi}=100$ GeV by including a nuclear form factor, but without recoil threshold effects.}
  \label{difrate}
\end{figure}

\begin{figure}
 \includegraphics[height=6.0cm,width=6.0cm]{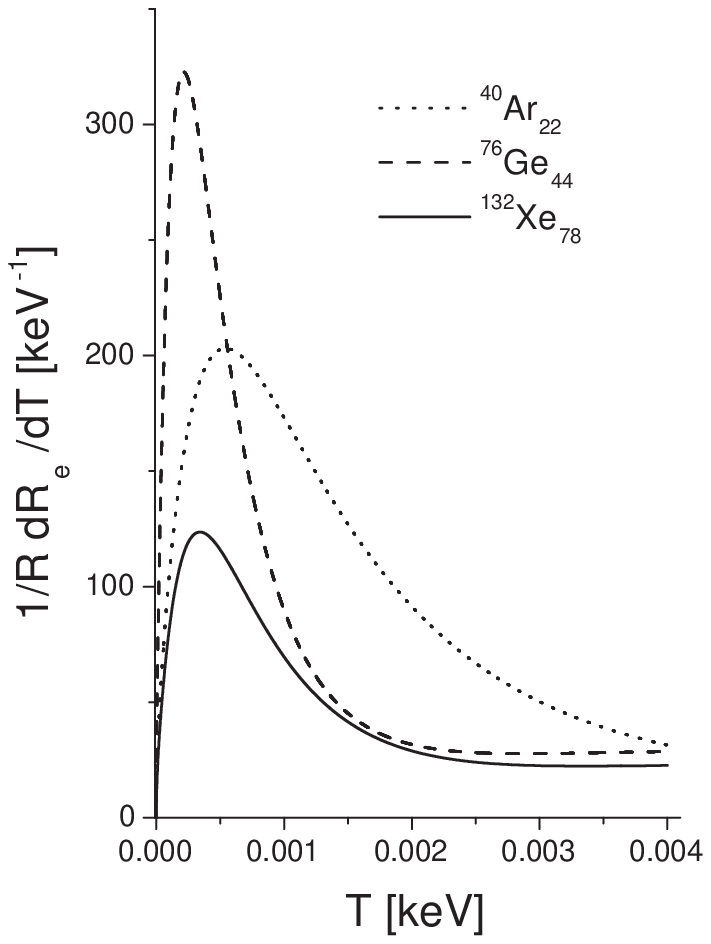}
\
\includegraphics[height=6.0cm,width=6.0cm]{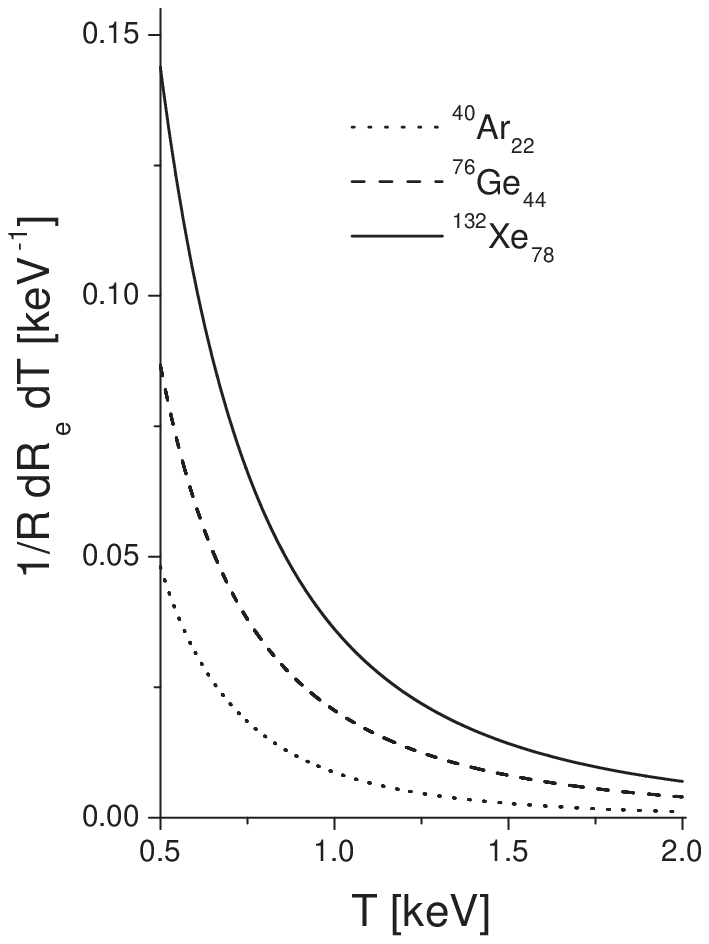}
\
\includegraphics[height=6.0cm,width=6.0cm]{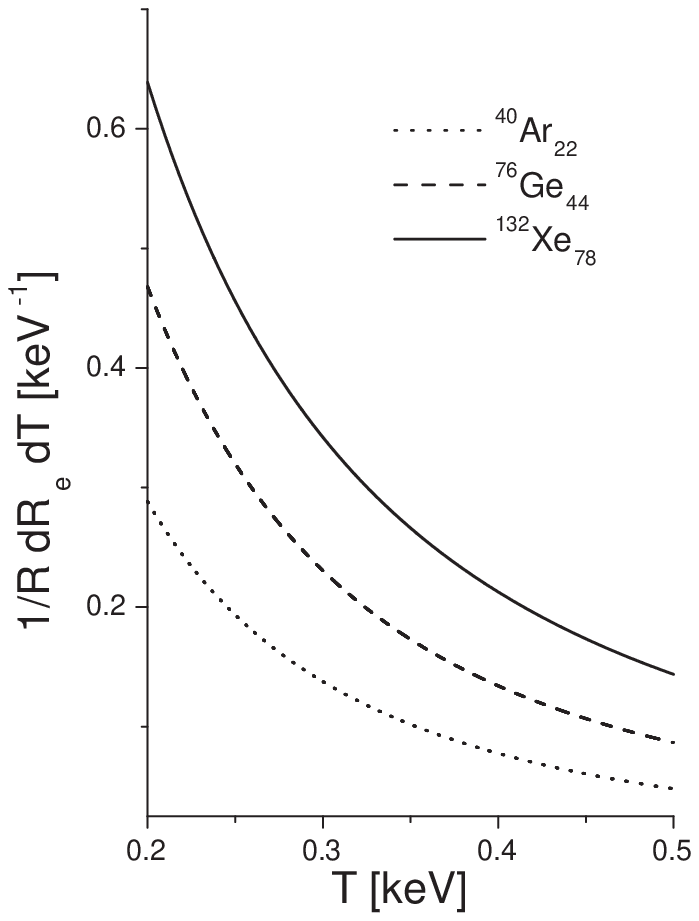}
 \label{difrateXe}
 \caption{The same as in Fig. \ref{difrate} except that
 we have now separated the plot in various
energy intervals, so that the energy dependence at higher energies
becomes visible.}
\end{figure}

\begin{figure}
 \includegraphics[height=8.0cm,width=8.0cm]{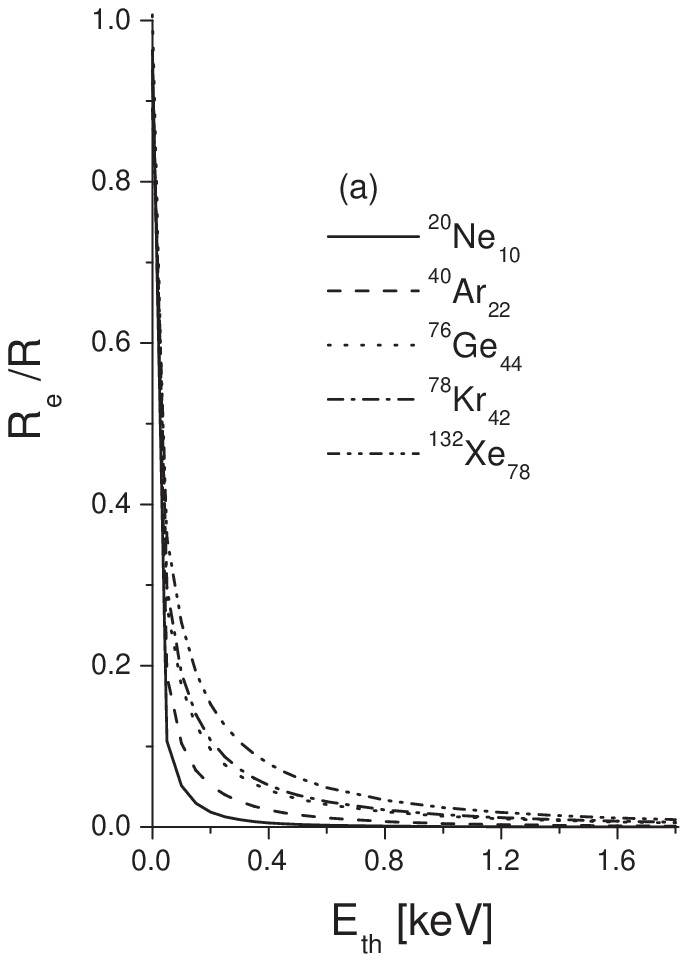}
\
 \includegraphics[height=8.0cm,width=8.0cm]{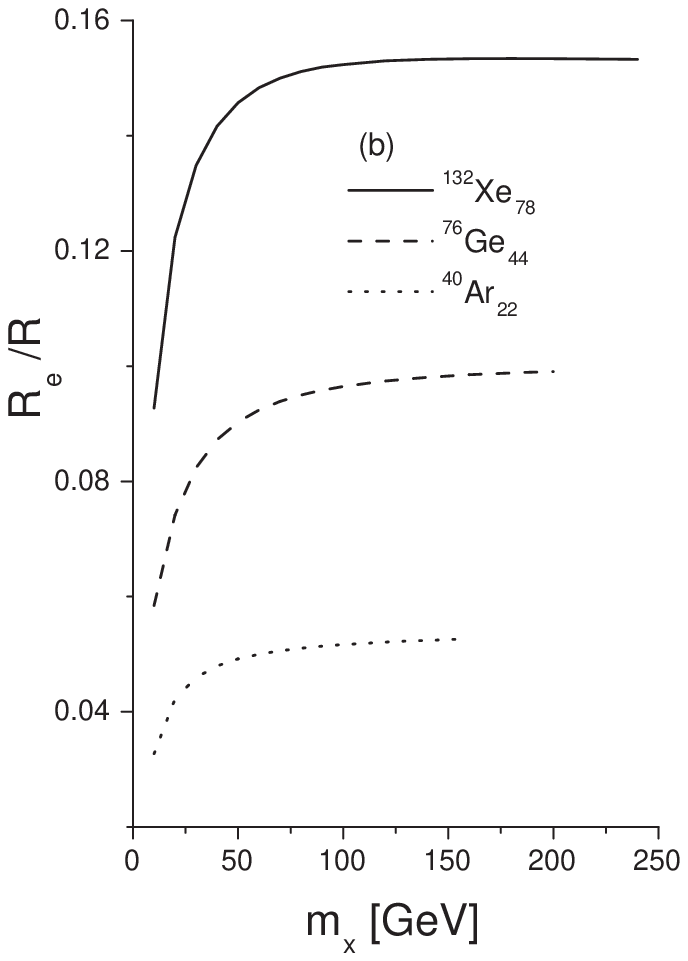}
 \caption{ (a) The total ionization rate per electron divided by the standard
 nuclear recoil rate as a function of the electron threshold energy. The results exhibited were obtained for
  a typical LSP mass $m_{\chi}=100$ GeV by including a
 nuclear form factor, but no threshold effects on recoils. (b) The same quantity plotted as a
function of the mass $m_{\chi}$ of the LSP. The results
 were obtained for a  threshold energy $E_{th}=0.2\ keV$}
 \label{totalrate0}
\end{figure}

\begin{figure}
 \centering
 \includegraphics[height=8.0cm,width=8.0cm]{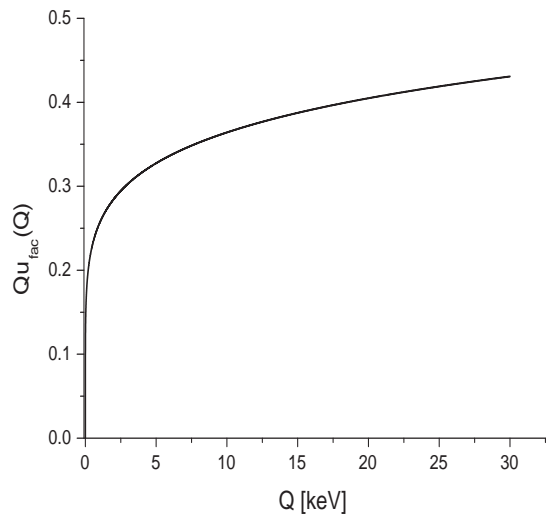}
  \caption{The quenching factor $Qu_{fac}(Q)$ as a function of the recoil energy Q in $keV$. }
\label{Qu.fig}
\end{figure}

\begin{figure}
 \includegraphics[height=12.0cm,width=12.0cm]{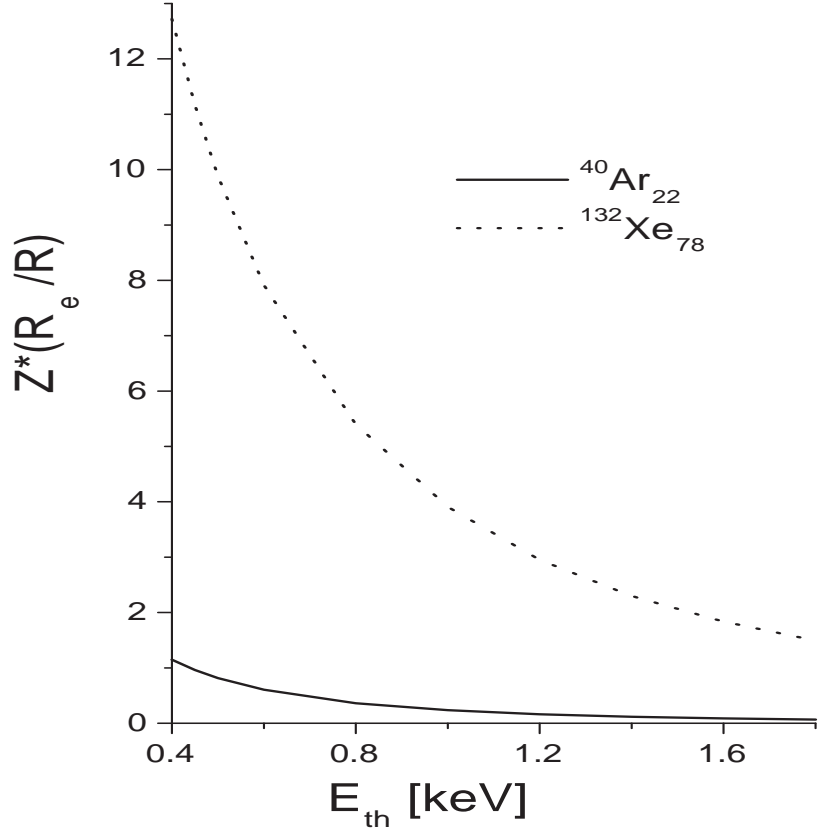}
 \caption{ The same as in Fig. \ref{totalrate0}a except that now the ionization rate $R_e$ per electron is
multiplied by the atomic number $Z$ to obtain the rate per atom.
In other words we plot the relative ionization rate $ZR_e$ per
atom with respect to the nuclear recoil rate, i.e. $r=Z
\frac{R_e}{R}$.
 The quenching factor of about $1/3$ has also been included in the denominator. More precisely
 we use (1/3) R for the denominator (standard recoil)  to
 incorporate the reduction of the  recoil rate with a typical threshold and
quenching factor (see Fig. \ref{Qu.fig}).
} \label{zefig}.
\end{figure}

\begin{figure}
 \includegraphics[height=8.0cm,width=8.0cm]{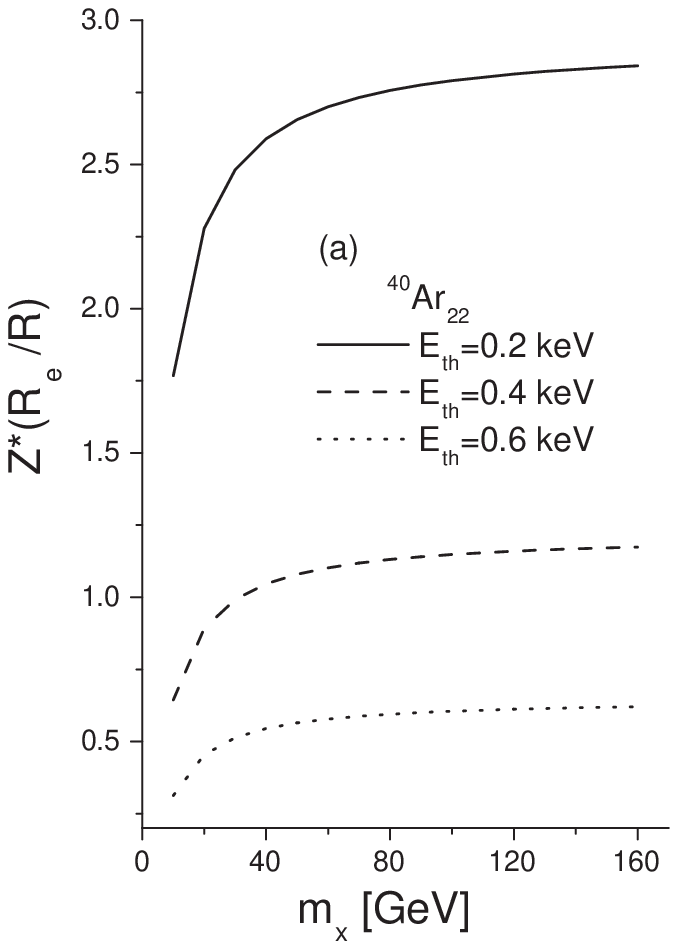}
\
 \includegraphics[height=8.0cm,width=8.0cm]{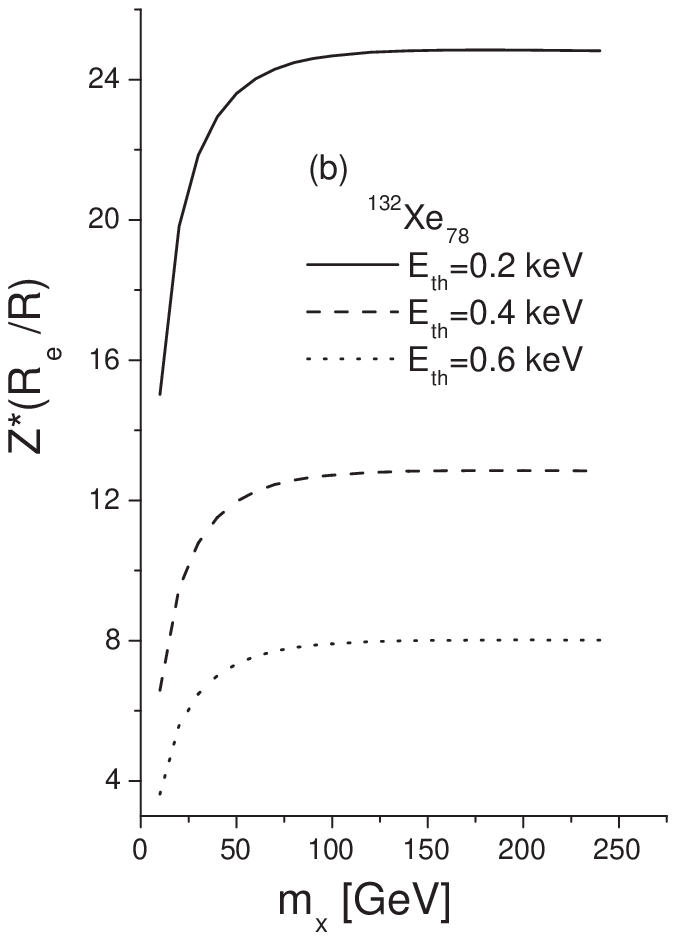}
 \caption{ The same as in Fig. \ref{totalrate0}b
except that now the electron ionization rate per electron has been
multiplied
 by the atomic number $Z$ (the relative event rate per atom is $r=Z\frac{R_e}{R}$).
 The quenching factor of about $1/3$ has also been included in the denominator (standard nuclear recoil).
} \label{zmxfig}.
\end{figure}

We begin our discussion by focusing on the differential rate. It
is useful to divide it by the total nuclear recoil rate, i.e.
$\frac{1}{R}\frac{dR_e}{dT}$, in order to make our results
independent of the details of the SUSY parameter space and related
uncertainties. From previous work \cite{VerEj} on $^{20}$Ne with
hydrogenic wave functions we know that the differential rate
sharply picks at low electron energies. This result is confirmed
by our present calculation employing realistic electron wave
functions. This is exhibited by considering three different atoms
($^{40}$Ar, $^{76}$Ge and $^{132}$Xe) for the typical LSP mass
$m_{\chi}=100$ GeV (see Figs. \ref{difrate}-\ref{difrateXe}). If
we then integrate the differential rate
$\frac{1}{R}\frac{dR_e}{dT}$ with respect to the electron kinetic
energy $T$, we obtain the relevant event rate ratio
$\frac{R_e}{R}$. The thus obtained results for a number of nuclei
are
 shown in Fig. \ref{totalrate0}a. They are presented
as a function of the threshold energy for electron detection
$E_{th}$ for LSP mass $m_{\chi}=100$ GeV. We also present them
 as functions of the LSP mass $m_{\chi}$, for $E_{th}=0.2$ keV
 (Fig. \ref{totalrate0}b).
 We clearly see that the results are very sensitive
 to the threshold energy. It is also clear that the heavier
 targets are favored. It is encouraging that branching
 ratios per electron of $10\%$ are possible, if one can reach threshold
 energies as low as $200$ eV, which is feasible for gas
targets \cite{VerEj}. It should be mentioned that in obtaining
these results we assumed threshold effects for the ionization
electrons, but not for the nuclear recoils. The nuclear recoil
rate is
less sensitive to the threshold in the sub-keV region.\\
 As is well known the recoil experiments suffer from the effect of quenching.
 The idea of quenching is introduced, since,
for low energy recoils, only a fraction of the total deposited
energy goes into
 ionization. The ratio of the amount of ionization induced in the gas
 due to nuclear recoil to the amount of ionization
 induced
by an electron of the same kinetic energy is referred to as a
quenching factor $Qu_{fac}(Q)$. This factor depends mainly on the
detector material, the recoiling energy $Q$ as well as the process
considered \cite{SIMON03}.
 In our estimate of $Qu_{fac}(Q)$ we assumed a quenching factor
 of the following empirical form motivated by the Lidhard
theory \cite{SIMON03}-\cite{LIDHART}:
\begin{equation}
Q_{fac}(Q)=r_1\left[ \frac{Q}{1keV}\right]^{r_2},~~r_1\simeq
0.256~~,~~r_2\simeq 0.153, \label{quench1}
\end{equation}
where $Q$ is the nuclear recoil energy. The quenching factor
$Qu_{fac}(Q)$ is shown in Fig. \ref{Qu.fig}. Since, as we have
mentioned, we expect a threshold energy of $10$ keV and the
quenching factor becomes $\approx0.3$ above this energy, the
branching ratio will increase by another factor of three, if the
quenching is taken into account.\\
 The number of electron events divided by the nuclear recoil
 events is obtained by multiplying the above
 branching ratios with the atomic number $Z$. This enhancement
 makes it possible to detect electron events
 not only with TPC detectors but nuclear double beta decay
 detectors \cite{DBD}, sensitive to detection of low energy
 electrons $\sim 1 keV$.\\
 With this number of events one wonders, if one can detect
 X rays formed after the inner shell
electron hole, created by the LSP- nucleus collision, is filled by
one of the outer electrons, with or without coincidence
measurement of the outgoing particles (electrons and/or recoiling
nuclei). In fact we find that some inner shell orbits, in
particular 2p and 3d, can make a
 sizable contribution. It thus seems that such
 X rays can be exploited. Discussion of this phenomenon as well as
 the background of the Auger electrons
involving the inner shell electrons \cite{VerEj} will be discussed
in detail in a separate publication.\\
So far, the ratio of the electron rate to the recoil rate is
normalized to one electron per atom.
 In practice, the total electron rate for given detector mass
 is obtained by multiplying the above rate per electron
by the atomic number $Z$.  Then the ratio of the total electron
rate to the recoil rate is plotted as a
 function of the threshold energy in Figs  \ref{zefig},  \ref{zmxfig}.
 Here in addition a quenching factor of $\approx 0.3$ is used for the nuclear recoil.
The thus obtained ratio is very impressive, even at an electron
threshold higher than $0.5$ keV. Thus it is interesting to search
for
 electrons with the high-sensitivity low-threshold detectors such as
 those used for double beta decays discussed
in recent review papers \cite{DBD}

\section{ Concluding Remarks}
\label{sec4} We have explored a novel process for the direct
detection of dark matter, that is by exclusive observation of the
low energy ionization electrons produced during the LSP-nucleus
collision. We have applied our formalism in a variety of atoms
employing realistic electron wave functions and nuclear form
factors. We have focused in noble gases, since we expect that low
energy electrons can better be detected by gaseous TPC detectors.
Our results, however, can be applied, if necessary,  in other
targets, e.g. those employed in neutrinoless double beta decay,
since we find that the details of the electronic structure are not
very
important. \\
To minimize the uncertainties arising from SUSY and to make the
presentation as simple as possible, we have calculated the ratio
of the rate associated with this process divided by the rate for
the standard nuclear recoils. We find that this ratio  per
electron slightly increases
 with the atomic number. It
depends appreciably of the electron and the nuclear recoil energy
cut offs. It is quite important to make the electron detection
energy cut off as low as possible. Branching ratios (per electron)
of about $10\%$ can realistically be expected for reasonable
choices of these threshold energies. Thus the ratio of the number
of electron events per atom divided by the number of recoils,
obtained from the previous ratio after multiplication with the
atomic number $Z$, can even become larger than unity. Thus, since
the ionization electrons extend to the keV region, their
contribution has to be taken into account in the the standard
nuclear recoil experiments as well. \\
 As it has been previously shown \cite{VerEj} the background problem are no worse than
those appearing in the standard recoil experiments. They can be
further reduced by coincidence experiments, by measuring the
ionization electrons in coincidence nuclear recoils. Finally it
should noted that the inner
 shell electron holes, created during the
LSP-nuclear collision, can be filled by emitting X rays and/or
Auger electrons. Such issues, which are of
experimental interest, will be discussed in detail elsewhere.\\
 In view of the above considerations it is realistic to search for
 WIMPs by measuring ionization electrons.

\section{Acknowledgments}
The first author (Ch.C. M.) acknowledges support by the Greek
State Grants Foundation (IKY) under contract (515/2005) and would
like to thank Prof. G.J. Gounaris for useful discussions
concerning the atomic wave functions and the electron binding
energies. The second author (J.D. V.) is indebted to the Greek
Scholarship Foundation (IKYDA) for support and to Professor
Faessler for his hospitality in Tuebingen.  The other author ( H.
E) thanks the Institute for Nuclear Theory at the University of
Washington and the Department of Energy for partial support during
the completion of this work. Finally all authors would like to
thank Prof. C.F. Bunge for kindly providing the data for the
atomic wave functions.

\newpage
\clearpage \clearpage
\clearpage
\clearpage


\clearpage

\clearpage

\clearpage
\end{document}